%% file: conference_101719.tex
\documentclass[conference]{IEEEtran}
\IEEEoverridecommandlockouts
\usepackage{cite}
\usepackage{amsmath,amssymb,amsfonts}
\usepackage{algorithmic}
\usepackage{graphicx}
\usepackage{textcomp}
\usepackage{xcolor}
\usepackage{soul}
\usepackage{hyperref}
\usepackage{todonotes}
\usepackage{multirow}
\usepackage{nccmath}
\usepackage{mathtools}
\usepackage{hyperref} 
\usepackage{setspace}

\usepackage[T1]{fontenc}
\usepackage[scaled=0.85]{beramono}
\usepackage{listings}
\usepackage{xspace}
\newcommand{\AlgName}{{FABULA}\xspace}

\def\BibTeX{{\rm B\kern-.05em{\sc i\kern-.025em b}\kern-.08em
    T\kern-.1667em\lower.7ex\hbox{E}\kern-.125emX}}

\makeatletter \def\@IEEEpubidpullup{8\baselineskip} \makeatother 

\usepackage{fancyhdr}
\usepackage{kantlipsum}
\fancypagestyle{plain}{
\fancyhf{}
\fancyhead[C]{Conference on \LaTeX} %% C or L or R. %\fancyfoot[L]{This is a notice}% %% C or L or R. \renewcommand{\footrulewidth}{0pt} %\renewcommand{\headrulewidth}{0pt} 
} \usepackage{eso-pic}

\begin{document}
\AddToShipoutPictureBG*{
\AtPageUpperLeft{
\setlength\unitlength{1in}
\hspace*{\dimexpr0.48\paperwidth\relax} \makebox(0,-0.75)[c]{\textbf{2023 IEEE/ACM International Conference on Advances in Social Networks Analysis and Mining (ASONAM)}}}}  

\IEEEoverridecommandlockouts
\IEEEpubid{
\parbox{\columnwidth}{\vspace{-4\baselineskip} Permission to make digital or hard copies of all or part of this work for personal or classroom use is granted without fee provided that copies are not made or distributed for profit or commercial advantage and that copies bear this notice and the full citation on the first page. Copyrights for components of this work owned by others than the author(s) must be honored. Abstracting with credit is permitted. To copy otherwise, or republish, to post on servers or to redistribute to lists, requires prior specific permission and/or a fee. Request permissions from \href{mailto:permissions@acm.org}{permissions@acm.org}.\hfill\vspace{-0.8\baselineskip}\\ \begin{spacing}{1.2}
\small\textit{ASONAM '23}, November 6-9, 2023, Kusadasi, Turkey \\
\copyright\space2023 Association for Computing Machinery. \\
ACM ISBN 979-8-4007-0409-3/23/11\ldots\$15.00 \\ \url{https://doi.org/10.1145/3625007.3627505}
\end{spacing}
\hfill}
\hspace{0.9\columnsep}\makebox[\columnwidth]{\hfill}}
\IEEEpubidadjcol

\title{FABULA: Intelligence Report Generation Using Retrieval-Augmented Narrative Construction\\

}

\author{\IEEEauthorblockN{Priyanka Ranade}
\IEEEauthorblockA{\textit{Department of CSEE} \\
\textit{University of Maryland, Baltimore County}\\
Baltimore, MD, USA \\
priyankaranade@umbc.edu}
\and
\IEEEauthorblockN{Anupam Joshi}
\IEEEauthorblockA{\textit{Department of CSEE} \\
\textit{University of Maryland, Baltimore County}\\
Baltimore, MD, USA \\
joshi@umbc.edu}
}
\maketitle

\begin{abstract}
Narrative construction is the process of representing disparate event information into a logical plot structure that models an end to end story. Intelligence analysis is an example of a domain that can benefit tremendously from narrative construction techniques, particularly in aiding analysts during the largely manual and costly process of synthesizing event information into comprehensive intelligence reports. Manual intelligence report generation is often prone to challenges such as integrating dynamic event information, writing fine-grained queries, and closing information gaps. This motivates the development of a system that retrieves and represents critical aspects of events in a form that aids in automatic generation of intelligence reports.

We introduce a Retrieval Augmented Generation (RAG) approach to augment prompting of an autoregressive decoder by retrieving structured information asserted in a knowledge graph to generate targeted information based on a narrative plot model. We apply our approach to the problem of neural intelligence report generation and introduce \AlgName, framework to augment intelligence analysis workflows using RAG.  An analyst can use \AlgName to query an Event Plot Graph (EPG) to retrieve relevant event plot points, which can be used to augment prompting of a Large Language Model (LLM) during intelligence report generation. Our evaluation studies show that the plot points included in the generated intelligence reports have high semantic relevance, high coherency, and low data redundancy.

%\AlgName's EPG stores event narrative plot points according to the Inverted Plot Pyramid (IPP) narrative structure, specifically developed for communicating news stories.   

%\AlgName's EPG stores event narrative plot points according to the Inverted Plot Pyramid (IPP) narrative structure, specifically developed for communicating news stories. We take advantage of a RAG approach to fine-tune the GPT-Neo  LLM to output an intelligence report given a set of narrative prompt sets extracted from the EPG. 

\end{abstract}


\section{Introduction}

The process by which information about critical events is disseminated, articulated, and shaped into news stories has greatly evolved since the proliferation of digital media and the World Wide Web. Intelligence analysts now take advantage of an abundance of online communication mediums to widely share and obtain reporting on a variety of critical events. Intelligence analysts rely heavily on manual techniques to extract evolving fine-grained event details over multiple Open Source Intelligence (OSINT) sources in \textit{real-time}. This evaluated information is then manually documented within an {intelligence report}, which is used during tactical operations by decision makers to manage, evaluate, and keep updated on evolving {events}, such as breaking news occurrences.

Intelligence reports are inherently structured to communicate \textit{narratives}, which are accounts of \textit{interconnected event incidents and actors} (plot points) evolving through some notion of \textit{time}. In journalism and storytelling, there have been several types of \textit{narrative plot structures} proposed to organize and convey event information. One of the most well suited for intelligence analysis is the Inverted Plot Pyramid (IPP) narrative structure (Figure \ref{fig:ipp}, Section \ref{narrtheory}) which is designed specifically for conveying news event details. It is becoming more clear that the intelligence analysis domain can benefit tremendously from techniques in \textit{computational narrative construction}, which utilize existing Information Retrieval (IR) methods such as document collection, query processing, and ranking to aid end users in comprehending disparate event information. Specifically, integrating narrative construction tasks to intelligence analysis workflows can alleviate the costly nature of intelligence report generation in three primary ways: (a) Automatically extracting event information from a dense collection of documents based on a schema, (b) Aiding in information triage during intelligence report generation, (c) Tracking and integrating evolving event information over time.

Recent advancements in Large Language Models (LLMs) have enabled state of the art results in automatic text generation tasks, presenting new opportunities in the computational narrative construction domain. For example, an end user can issue directed \textit{prompts} to a generative LLM to automatically generate summaries that communicate end to end narratives about an event. Similarly, we envision that an \textit{intelligence analyst} can prompt an LLM to automatically generate accurate intelligence reports about queried events. Despite these potential benefits, a direct application of LLMs for automatic intelligence report generation presents several limitations, such as: (a) Output hallucinations where the generated text contains non-factual, non-event related, and incomplete information, (b) Lack of provenance, attribution, and trust for knowledge sources used to generate responses. The AI community has started to address these challenges through a general approach called \textit{Retrieval Augmented Generation} (RAG) which uses non-parametric memory to augment LLM generation. 

Inspired by these recent developments, we develop \AlgName, a framework that integrates a novel RAG approach for using narrative plot structures, LLMs, and knowledge graphs to automatically generate intelligence reports (Figure \ref{fig:sysarch}). \noindent The \textit{main contributions} of this paper include: 
\begin{itemize}
\item \AlgName: A framework to augment intelligence analysis workflows. Analysts can use the system to automatically generate intelligence reports for events utilizing contextual narrative features found in OSINT (Section \ref{systemdescription}).
\item Retrieval-Augmented Generation (RAG) approach which retrieves plot points from a knowledge graph and provides them as input prompt sets for guided LLM intelligence report generation (Section \ref{section:rag}).
\end{itemize}

\section{Related Work}

In this section, we describe research on narratives, news event OSINT, knowledge representation, and data to text generation. 

\subsection{Narratives, Stories, \& News}
\label{narrtheory}
OSINT about \textit{events}, are published via blogs, social networks, news sources. Events contain \textit{plot points}, which are incidents that directly impact what happens next \cite{lehnert1981plot}. 
Events are communicated through the form of \textit{narratives} which are accounts of interconnected plot points \cite{ranade2022computational}.  
A seminal example of a narrative plot structure is the \textit{Plot Pyramid Model} by Gustav Freytag, a five component framework that outlines thematic and temporal stages in generic storytelling \cite{boyd2020narrative}. The components are, \textit{Introduction, Rising Action, Climax, Falling Action, and Denouement}. The plot points in the Freytag pyramid develop and conclude progressively over time, first leading to the development of the \textit{climax} (introduction and rising action) and successively concluding to the \textit{denouement} as a direct result of the climax (falling action). Unlike the pyramid, other narrative plot structures organize plot points in varying ways. For example, the \textit{Fichtean Curve}, begins immediately with the rising action component, followed by a series of \textit{crisis} (falling action) \cite{thompson1999storytelling}. There are several more examples of other narrative plot structures such \textit{The Hero's Journey} that model the development of events differently % through varying components 
\cite{thompson1999storytelling}. 
Our system \AlgName, focuses specifically on modeling the plot structure of open source \textit{news events}. A narrative plot structure specifically developed for communicating news stories is the \textit{Inverted Plot Pyramid} (IPP) (Figure \ref{fig:ipp}). IPP is a three component model that conveys the critical plot points in the first component \textit{(Lead)}, the event developments \textit{(Body)}, and the nonobligatory information at the end \textit{(Tail)} \cite{keith2020evaluating}. 

%Though narratives are commonly employed methods for connecting plot points of an event in traditional storytelling, there is a gap in the Information Retrieval (IR) and Artificial Intelligence (AI) literature for utilizing narrative plot structures during the information retrieval of disparately sourced news events. In the next section, we describe existing methods in event ordering in relation to \textit{narrative construction}.

%Rather than constructing event details together based on \textit{plot context}, current methods focus on event ordering technologies based on fine-grained textual information such as timestamps and causal semantic information. In the next section, we describe existing methods in event ordering in relation to \textit{narrative construction}.

\subsection{Narrative Construction and Schemas} 
Sequencing disparate events from a variety of sources is known as \textit{fragmented narrative construction} \cite{ranade2022computational}. While the disparate nature of OSINT provides opportunities for users to obtain insights, it presents challenges for chaining accurate data across \textit{noisy} sources \cite{el2020supporting}. There have been several methods that address this problem.
One is by sequencing causal and temporal event shifts into \textit{story chains}. Zhu et al. \cite{zhu2012finding} defines a story chain as ``a construction of news articles that reveal hidden relationships among different events''. They utilize random walks on a bipartite graph to form a coherent story chain based on a query. Prior to this work, research projects have ordered news based on hierarchy\cite{shahaf2010connecting,fung2007time}. A novel method utilized in our \AlgName system for extracting narratives is through the use of narrative schemas. Narrative schemas are models used to represent primary components of a narrative, such as actors, plot points, and actions \cite {meghini2021representing}. 

%Yan et al. create functional story schemas that capture latent, structural features in \textit{Reddit} conversation threads \cite{yan2019using}. Similarly, Labutut et al. model actor relationships using character networks \cite{labatut2019extraction}.

\subsection{Information Retrieval \& Knowledge Representation}

Information Retrieval (IR) systems have evolved from symbolic-based methods \cite{singhal2017pivoted} to neural retrieval models \cite{craswell2018neural}. These capture \textit{semantic matches} using neural networks to build vectorized knowledge representations. Representation techniques such as Knowledge Graphs (KG), have been popularly employed to \textit{support} IR tasks \cite{dietz2018utilizing}. 

%In \AlgName, we utilize a KG and a base event narrative ontology to retrieve news narrative plot points for intelligence report generation. 

%Lately, KGs have been used as input to Large Language Models (LLM) for text generation. In the next section we review approaches that combine KGs and LLMs for downstream tasks such as generation. Our system \AlgName augments the LLM generation process using KG aided prompting. 

%In this paper, we utilize ontological class hierarchies and instance graph data as input to decoder large language models to better construct event details (plot points) together into readable narratives. 

\subsection{Knowledge Graph (KG) to Text Generation}

The KG-to-text generation task is a form of semantic triple verbalization, which automatically generates descriptive text for a given KG \cite{koncel2019text, ke2021jointgt}. State of the art methods fine-tune text-to-text and generative decoder pre-trained models with KG-to-text datasets. One approach to this problem is \textit{Retrieval Augmented Generation} (RAG), a method that aims to seed external data is a technique that uses retrieved data that is stored externally (like in a KG) from the foundation model, which is used to augment LLM prompting by injecting relevant retrieved information.

%Though such methods have shown impressive capabilities in generating fluent text from KG triples, some limitations include the inability to encode graph structure during training and issues in explicitly modelling graph-text alignments \cite{ke2021jointgt}. 
%We utilize a KG as an external store in \AlgName. Utilizing KGs as the external data store in RAG-based set ups continue to emerge, particularly for question answering tasks \cite{kang2023knowledge,sha2023retrieval}.

%\section{Problem Statement and Methodology}

\section{Methodology}
\label{systemdescription}

\begin{figure}[hb]
    \centering
    \vspace{-3mm}
\includegraphics[scale=0.4]{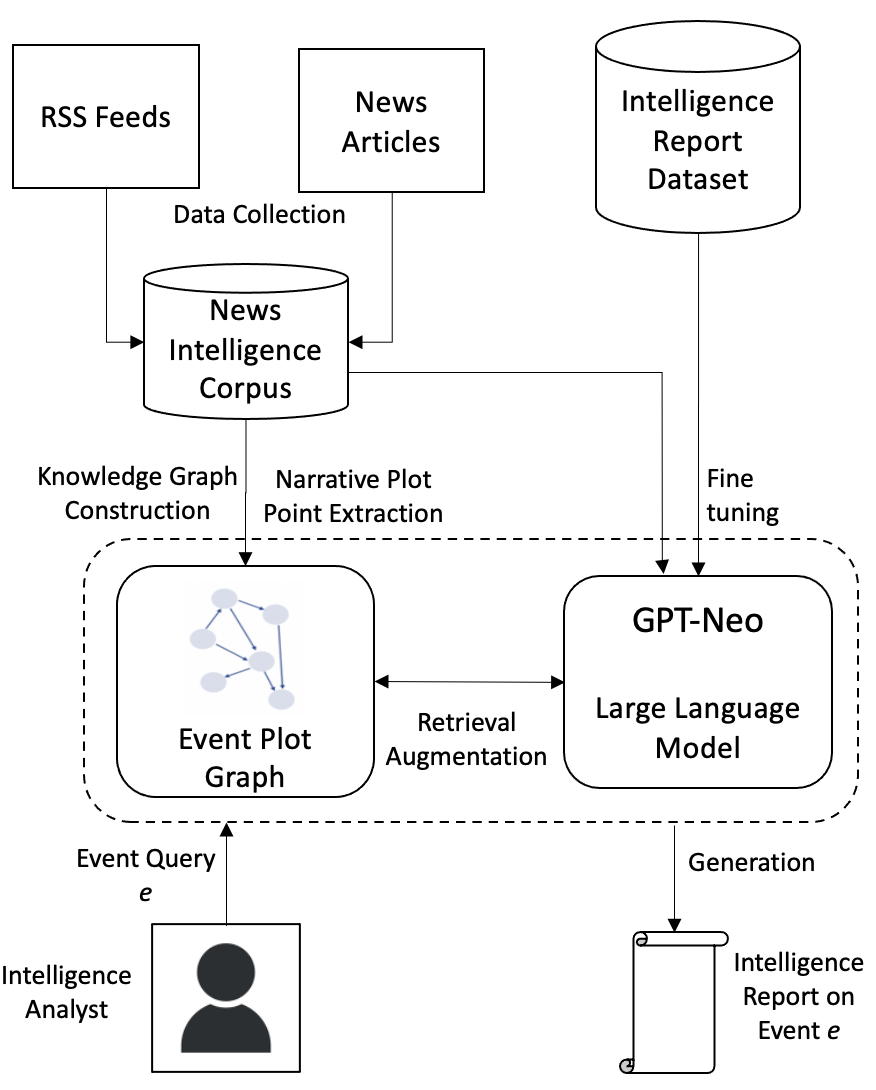}
\vspace{-2mm}
    \caption{\AlgName System Architecture and Data Flow.}
    \label{fig:sysarch}
    \vspace{-2mm}
\end{figure}

This work is guided by the following research question: 
\textit{Does incorporating narrative-based features during Retrieval-Augmented Generation (RAG) produce intelligence reports with high semantic relevance, high event coherency, and little to no hallucination?} Our approach is applied and evaluated specifically in the context of narrative construction for news events represented in open sourced news articles.

Suppose we have a set of randomly ordered articles $d_{1}$, $d_{2}$..., $d_{n}$, retrieved by a keyword search query about an event $e$, which contain several plot points $p_{1}$, $p_{2}$..., $p_{n}$. Each plot point is extracted, ranked, and ordered into a chronological sequence. Consider a real-world example where an \textit{intelligence analyst} requires information about a critical event $e$. The analyst will input a query about $e$, retrieve relevant information from a multiple set of sources (news blogs, social media posts) to write a condensed intelligence report. We develop \AlgName, an analyst-augmentation framework that integrates real-time news event retrieval, narrative schema-based information extraction and representation of event concept information, and retrieval-augmented generation (RAG) of intelligence reports.

Our approach is displayed in Figure \ref{fig:sysarch}. We begin by creating a \textit{news intelligence corpus} from popular news sources and publicly available U.S Intelligence Community (IC) reports, described in Section \ref{section:corpus}. This corpus is further condensed through information extraction of plot points contained within news articles and follows the Inverted Plot Pyramid (IPP) narrative structure (Section \ref{section:plotextraction}). 
The extracted plot points are then asserted into an Event Plot Graph (EPG) using our base Event Narrative Ontology (ENO) schema (Section \ref{section:eno}). Next, we fine-tune the LLM, GPT-Neo \cite{black2021gpt} using our news intelligence corpus and the extracted plot points \ref{section:finetuning}). \AlgName's EPG is queried using Retrieval Augmented Generation (RAG) to control GPT-Neo's intelligence report generation process. 

\begin{figure}[ht]
    \centering
    \vspace{-3mm}
\includegraphics[scale=0.27]{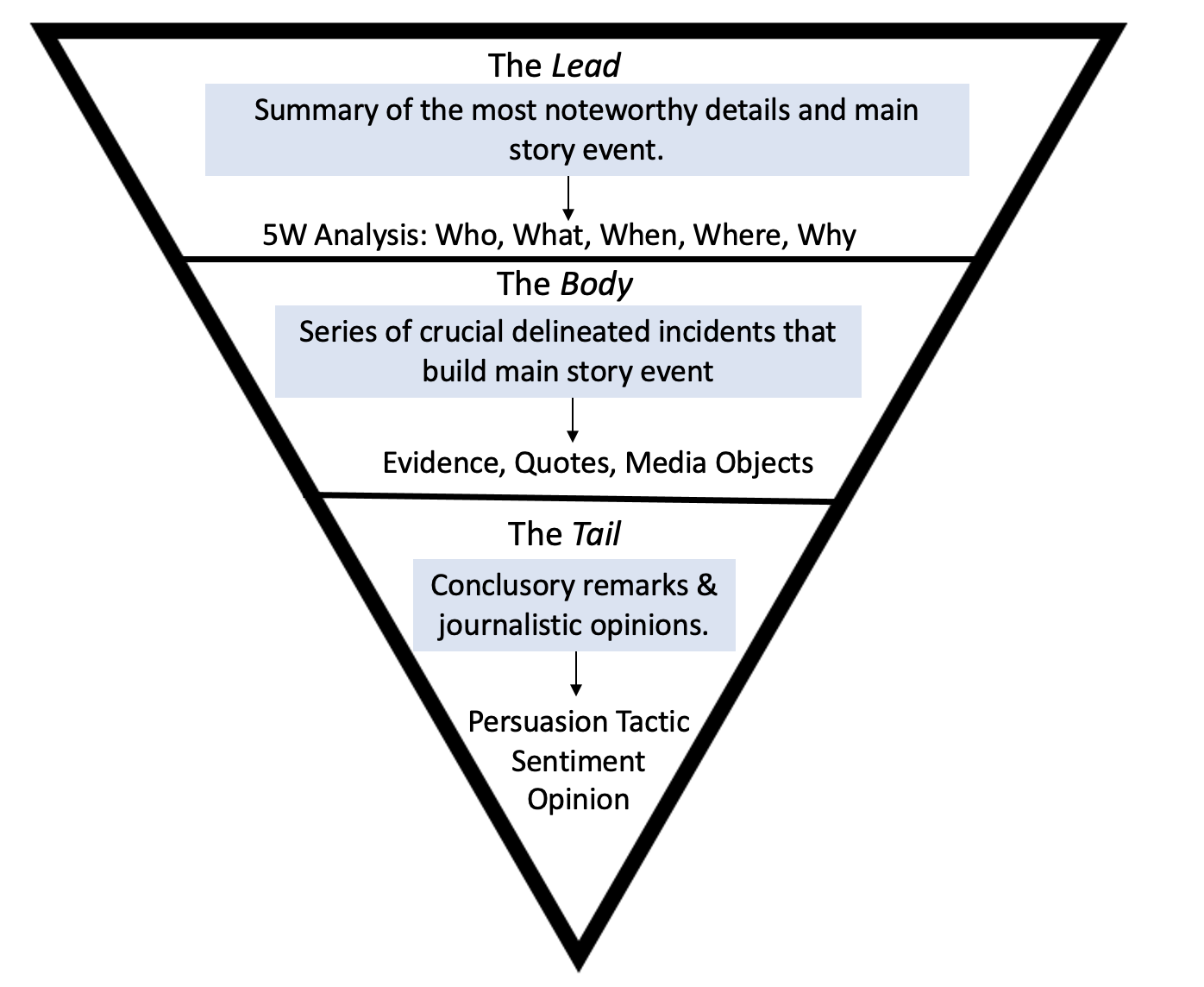}
\vspace{-2mm}
    \caption{The Inverted Pyramid Plot (IPP) model and associated text features.}
    \label{fig:ipp}
    \vspace{-5mm}
\end{figure}

\input{plottable}
\section{Event Intelligence Collection and Representation}

In this section we describe methods for collecting open source news event data and organizing it in a knowledge representation based on the Inverted Pyramid Plot (IPP) news narrative model features.

\subsection{News Intelligence Corpus}
\label{section:corpus}
The first component of \AlgName is a \textit{news intelligence corpus} which contains streams of scraped public news articles, $D$ and open source intelligence reports $IR$ released by the U.S government. \AlgName utilizes Really Simple Syndication (RSS) feed triggers for sources such as CNN \cite{cnn}, New York Times \cite{nytimes}, CBS News \cite{cbs}, U.S. Department of State \cite{usdos}, and U.S. Department of Defense \cite{usdod}. Each feed contains article metadata such as headline, author(s), abstract, and a web link. When updates are pushed by sources on their RSS feeds, \AlgName utilizes the RSS web links to retrieve the corresponding webpages. We further extract information such as timestamps, news text, image links, video links from each article, using the BeautifulSoup web parser\footnote{\url{https://www.crummy.com/software/BeautifulSoup/}}. Each individual news article is represented in the collection of OSINT news set $D$.
\[D = \{d_1, d_2, ...,d_k, ..., d_n\}\]

Each $d_k$ represents a news article and its headlines, author(s), timestamps, text, images, video links.

Set $IR$ is composed of publicly available intelligence reports released by the United States Office of the Director of National Intelligence (ODNI) \footnote{\url{https://www.dni.gov/index.php/newsroom/reports-publications}}. The ODNI intelligence reports (2005-2023) have a common structure that matches the Inverted Plot Pyramid narrative scheme (Section \ref{section:plotextraction}). Our final news intelligence corpus $D+IR$, contains to 3000 $D$ news articles and 165 $IR$ reports.

\subsection{Event Plot Extraction} 
\label{section:plotextraction}
\AlgName solves a \textit{fragmented narrative construction} problem where 
given a set of relevant articles %$RD = \{rd_1, rd_2, ...,rd_l, ..., rd_m\}$, $RD \, \epsilon \, D$, 
retrieved by a search query about event $e$, contain several plot points:
\[P = \{p_1, p_2, ..., p_i, ..., p_o\}\]
Where, $p_i$ is a plot point that is extracted from $D$. The associated plot points for an event $e$, can be determined through information extraction % of underlying narrative elements 
based on classes %that compose 
in the \textit{Inverted Plot Pyramid (IPP)} (Figure \ref{fig:ipp}), a standardized narrative structure for communicating news events (Section \ref{narrtheory}). The IPP narrative structure has three sub-categories: \textit{Lead, Body,} and \textit{Tail}. Definitions of the IPP sub-categories and associated sub-types are available in {Table} \ref{tab:ipp}. To extract the IPP-based plot points, we implement a \textit{Narrative Plot Concept Extractor (NPCE)} that processes the set of news articles $D$. In the rest of this section, we further describe the methods used to extract the plot points described in {Table} \ref{tab:ipp} that form set $P$. Extraction occurs at each level of the IPP - \textit{Lead, Body,} and \textit{Tail}.

\begin{table}[ht]
\footnotesize
\centering
\begin{tabular}{|l|l|} 
\hline
\textbf{Lead Class Attributes}~ & \textbf{OntoNotes Entity and Relationship~Types}                                                                                                   \\ 
\hline
Who                    & PERSON,~NORP,~ORG~ ~~                                                                                                                     \\ 
\hline
What                   & \begin{tabular}[c]{@{}l@{}}EVENT, FAC,~PRODUCT,~WORK\_OF\_ART,\\LAW,~MONEY,~LANGUAGE, PERCENT,\\QUANTITY, ORDINAL, CARDINAL\end{tabular}  \\ 
\hline
When                   & DATE, TIME                                                                                                                                \\ 
\hline
Where                  & GPE,~LOC                                                                                                                                  \\ 
\hline
Why                    & \begin{tabular}[c]{@{}l@{}}cause, causing, caused by,\\because,~since,~after,~for,~as~and~of~ ~ ~~\end{tabular}                           \\
\hline
\end{tabular}
\vspace{2mm}
\caption{Entity and Relationship types for the 5W classes}
\label{tab:5w}
\vspace{-6mm}
\end{table}
%\todo{caption}

\subsubsection{The Lead}
Event information contained in the IPP {Lead} class describes the most noteworthy event details. As indicated in Table \ref{tab:ipp}, these details are expressed through the \textit{5W communication device}: who? ($p_{who}$), what? ($p_{what}$), when? ($p_{when}$), where? ($p_{where}$), and why? ($p_{why}$). The answers to the 5W questions provide a circumstantial view of an event. The NPCE extracts the 5Ws using pre-trained Named Entity Recognition (NER) and Part of Speech (POS) Tagging models from the standardized spaCy NLP Framework \cite{spacy}. The specific entity and relationship types extracted are provided in Table \ref{tab:5w}. The spaCy NER was trained on the widely benchmarked OntoNotes dataset. \cite{schmitt2019replicable}. The \textit{why} category in particular, is extracted using the spaCy Part of Speech (POS) tagger, which locates sentences containing causal relationships (prepositions, verbs, conjunctions) between entities.

\subsubsection{The Body}
The most significant plot points, involving major incidents and themes of an event are communicated in the body of an article. The incidents are typically written as factual occurrences to form a delineated sequence of information that builds the main story objective and overall situational awareness of an event. \AlgName's NPCE utilizes a FAISS-based clustering \cite{jegou2022faiss} and regex approaches to extract Body category event information (Table \ref{tab:ipp}), which include evidences ($p_{evidence}$), quotes ($p_{quote}$), and media objects stored as URLs ($p_{photo}$, $p_{video}$, $p_{audio}$).

\subsubsection{The Tail}

The Tail category contains conclusive remarks, journalistic opinions ($p_{opinion}$), persuasion tactics ($p_{tactic}$), and perceived sentiment ($p_{sentiment}$) of the article author and source organization. These features do not contain plot points that impact the \textit{development} of an event, but rather can be used by an analyst to understand factors that may influence the author's narrative.

Opinion ($p_{opinion}$) and persuasion tactic ($p_{tactic}$) identification is a multi-label task at the paragraph level. We utilize a gold standard model 
developed for the ACL Semantic Evaluation Task \cite{piskorski2023semeval}. This was trained on the only publicly available human-labeled corpus specifically developed for persuasion language extraction \cite{piskorski2023semeval}. We treat each article in our set $D$ as a holdout sample and use the provided model to extract the persuasion tactics and the corresponding text sample. For sentiment detection, we utilize the spaCy sentiment analysis program \textit{polarity} to extract positive, negative, and neutral sentiment for each article. 

The NPCE output of the 3-level \textit{Lead, Body,} and \textit{Tail} extraction populates the event plot points set $P$. We maintain the mapping between the extracted set $P$ and the associated documents in set $D$ for LLM prompt tuning, described further in Section \ref{section:rag}. In the next subsection, we describe methods we use to assert set $P$ into \AlgName's Event Plot Graph (EPG).

\subsection{Event Narrative Ontology \& Event Plot Graph}
\label{section:eno}

We introduce the \textit{Event Narrative Ontology} (ENO), a Web Ontology Language (OWL)-based knowledge representation. ENO allows \AlgName to store event information based on extracted plot point category features, described in the previous section. 
%\textbf{This allows us to incorporate contextual ordering during Retrieval Augmented-Generation}.
ENO serves as the base schema for \AlgName's \textit{Event Plot Graph} (EPG). NPCE's output, set $P$ (Section \ref{section:plotextraction}) is asserted in the EPG using ENO. ENO classes and properties have been constructed using the elements of the \textit{Inverted Plot Pyramid (IPP)} narrative scheme (Figure \ref{fig:ipp}, Section \ref{narrtheory}). While we incorporate the IPP scheme due to its relevance to intelligence analysis in particular, variants of ENO can be built based on a variety of narrative theories.

The EPG contains a set of stored as Resource Description Framework (RDF) triples donated as $G$,
%\begin{fleqn}[\parindent]
\begin{equation} \label{eq11}
G = \{(s, p, o) | s, o \, \epsilon \, I, p \, \epsilon \, R \}
\end{equation} 
%\end{fleqn}
\noindent where, $I$ and $R$ denote the instances and relations stored in $G$. $(s, p, o)$ is a single triple in $G$ and denotes the relation $p$ between two entities $s$ and $o$. Below is a description of ENO classes and properties. 

\textit{Classes in ENO:} ENO contains a total of 16 classes and subclasses: two generic classes: $NewsArticle$, $PlotPoint$, and 14 subclasses, all of which are of type $owl:Class$. The classes organize the information extracted in Section \ref{section:plotextraction} in a form that incorporates the IPP narrative scheme. Descriptions are as follows:

\begin{itemize}
\item \texttt{NewsArticle}: Instances contain identifiers for news articles and metadata such as publisher, author, URL, etc. 
\item \texttt{PlotPoint}: Describes IPP narrative elements (Section \ref{section:plotextraction}). It has following subclasses: 
 %\begin{itemize}
  \subsubsection{\texttt{Lead}}: Plot points that include noteworthy details. Subclasses are categorized %based on the subclasses noted 
  below:
  \begin{itemize}
\item \texttt{Who}: Person, affiliation, organization ($p_{who}$).
  \item \texttt{What}: Incidents, artifacts, or actions ($p_{what}$). 
  \item \texttt{When}: Recorded timestamps and dates ($p_{when}$).
  \item \texttt{Where}: Geographic locations/regions ($p_{where}$).
  \item \texttt{Why}: Event causal descriptions %providing reasoning for an event occurrence 
  ($p_{why}$).
  \end{itemize}
  \subsubsection{\texttt{Body}}: Plot points that describe news article objective. It has the following subclasses:
  \begin{itemize}
    \item \texttt{Evidence} Supporting details %, background %information %that provides  and background for queried event 
    ($p_{evidence}$).
  \item \texttt{Quote}: Text demarcated by quotation marks %, noted by involved persons 
  ($p_{quote}$). 
  \item \texttt{MediaObject}: Class representing photo ($p_{photo}$), audio ($p_{audio}$), and video ($p_{video}$) DOM objects. 
  \end{itemize}
  \subsubsection{\texttt{Tail}}: Plot points representing closing remarks and opinions. It has the following subclasses: %\todo{class descriptions for tail}  
  \begin{itemize}
    \item \texttt{Opinion}: Extracted author opinion %statements of the news article author(s) 
    ($p_{opinion}$). 
    \item \texttt{PersTactic}: Persuasion technique ($p_{tactic}$).
    \item \texttt{Sentiment}: Tone, %affective state, 
    emotion, mood ($p_{sentiment}$). 
     \end{itemize}
 \end{itemize}

\noindent\textit{Properties in ENO:}
To encode extracted relations ENO incorporates multiple object and data properties that can be asserted in the EPG. We describe some of these below:
\begin{itemize}
\item $articleHeadline$: Extracted string literal from \textit{NewsArticle} instance denoting the article headline.
    \item $authorOfArticle$: Extracted string literal from \textit{NewsArticle} instance denoting author(s) of the article.
     \item $publishedBy$: Data property denoting the publishing source of the \textit{NewsArticle} instance. %This property enables the system to tract provenance of the information in \AlgName. 
     \item $publishedDate$: Timestamp for $NewsArticle$ instance.
    \item $hasPlotPoint$: This object property helps codify relations between instances of the extracted IPP $PlotPoint$ and its associated $NewsArticle$. The property can be inherited by instances of $PlotPoint$ subclasses.
\end{itemize}
\begin{figure}[h]
    \centering
    \vspace{-2mm}
    \includegraphics[scale=0.25]{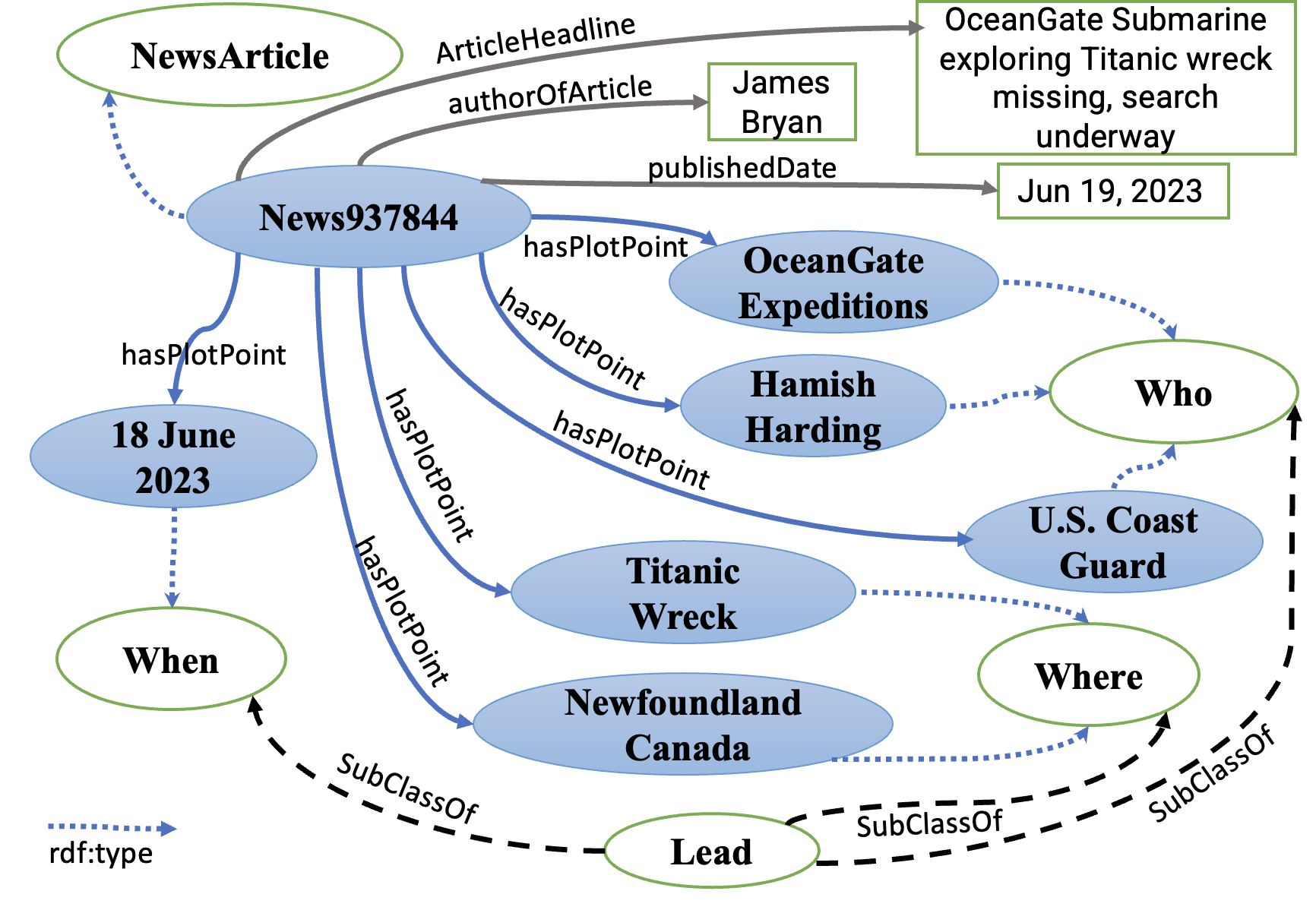}
    \vspace{-2mm}
    \caption{Populated EPG Sub-graph for the 2023 Titan Submersible Implosion.}
    \label{fig:graph}
    \vspace{-2mm}
\end{figure}
For example, Figure \ref{fig:graph}, shows an EPG sub-graph %graphical representation of intelligence extracted from multiple news reports from NPR, FoxNews, NBC, etc. 
about the \textit{2023 Titan Submersible Implosion} event. The graph represents one news article \textit{News937844}. %The output of the Narrative Plot Concept Extractor
%(NPCE) described in Section \ref{extract} is asserted in \AlgName's EPG using the ENO base schema. 
This sub-graph displays the \textit{Lead} narrative plot points about a \textit{catastrophic implosion} ($p_{what}$), involving the \textit{U.S Coast Guard}, \textit{OceanGate Expeditions}, and \textit{Hamish Harding} ($p_{who}$), occurring at the \textit{Titanic Wreck} site, near \textit{Newfoundland, Canada} ($p_{where}$).%that is further being investigated due to \textit{missing passengers} ($p_{why}$). 

Though not shown in this subgraph, we were able to extract \textit{Body} and \textit{Tail} narrative plot points, which were asserted in the EPG. These include \textit{Body} instances: four $p_{evidence}$ pieces, two $p_{quote}$ and one $p_{image}$, and \textit{Tail} instances: an \textit{attack on reputation} $p_{tactic}$ and \textit{negative} $p_{sentiment}$ utilized by the article author \textit{James Bryan}. 
%\todo{connect with table Pwhat} 
The EPG $G$ constructed from the news stream collection $D$ is next utilized in automatic intelligence report generation.

\section{Retrieval-Augmented Intelligence Report Generation}
\label{section:reportgen}

\begin{figure}[h]
    \centering
    \vspace{-3mm}
\includegraphics[scale=0.42]{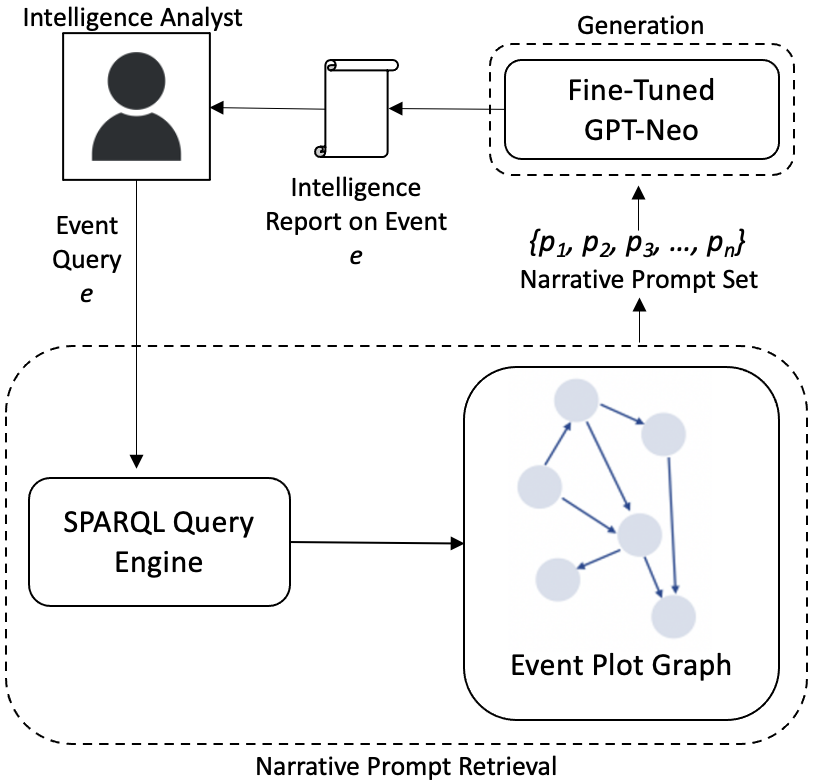}
    \vspace{-2mm}
    \label{fig:rag}
    \caption{\AlgName's Retrieval-Augmented Generation (RAG) of Intelligence Report about event $e$.}
    \vspace{-2mm}
\end{figure}

An intelligence analyst can use \AlgName to delineate event plot elements stored in the EPG $G$, for automatic generation of intelligence report $Y$ about event $e$. \AlgName implements a Retrieval-Augmented Generation (RAG) approach that queries the EPG to retrieve a \textit{narrative prompt set} about event $e$ (Figure \ref{fig:rag}). The set serves as a prompt to a Large Language Model (LLM) during report generation. There are three primary steps to our approach: (1) Fine-Tuning an LLM with $D+IR$, (2) Data-to-text prefix-tuning, and (3) SPARQL template LLM prompting for report generation. 

Let $V$ denote the vocabulary set of the report generation task. The desired \textit{target output} is to generate report text denoted by $Y$ utilizing the LLM where, 
\[Y = (w_1, w_2, ..., w_j, ..., w_T)\]
$w_j \, \epsilon \, V$ is a single word in the generated report $Y$. To generate $Y$, we first fine-tune the LLM GPT-Neo \cite{black2021gpt} with our news intelligence corpus $D+IR$. 
Fine-tuning GPT-Neo using $D+IR$ has two advantages. First, it augments the existing vocabulary of GPT-Neo ($V_{GPT-Neo}$) with the vocabulary of the news intelligence corpus ($V_{D+IR}$) which is equivalent to the vocabulary of \AlgName's EPG. After the fine-tuning report generation task, vocabulary $V$ includes both $V_{GPT-Neo}$ and $V_{D+IR}$. Second, fine-tuning with public intelligence reports $IR$ provides the LLM with examples of the desired document structure for output $Y$.

To generate the intelligence report $Y$, GPT-Neo takes as input a \textit{narrative prompt set}, i.e. a set of narrative plot points about the event $e$ stored in the subgraph $G' \, \epsilon \: G$. 
\[G' = \{(s_{e}, p, o_{e}) | s_{e}, o_{e} \, \epsilon \, I_{e}, p \, \epsilon \, R \}\]
\noindent where, $I_{e} \, \epsilon \, I$ and $R$ denote the instances relevant to query $e$ and relations stored in $G$ (Equation (1). The determination of the narrative plot points in $G'$ retrieved from $G$, is implemented using the SPARQL Protocol and RDF Query Language templates executed on \AlgName's EPG $G$. 

The narrative prompt set serves as input to the fine-tuned GPT-Neo LLM that outputs the intelligence report $Y$. 
Each component of our approach is further described in the rest of this section. 

\subsection{Fine-Tuning GPT-Neo}
\label{section:finetuning}
%In \AlgName, to generate the intelligence report $Y$ for a queried event $e$, we fine-tune GPT-Neo. 
Fine-tuning is an example of transfer learning, a method that seeds additional domain knowledge to a pre-trained Large Language Model, without training all parameters from scratch. We fine-tune the 1.3B parameter GPT-Neo decoder \cite{black2021gpt}. The original GPT-Neo model was trained with the \textit{Pile} dataset \cite{gao2020pile}, which is an 800GB English text corpus that consists of 22 high quality datasets. Fine-tuning GPT-Neo using $D+IR$ (Section \ref{section:corpus}) augments the existing vocabulary of GPT-Neo ($V_{GPT-Neo}$) with the vocabulary of the news intelligence corpus ($V_{D+IR}$) and allows GPT-Neo to model the format and syntactic style of known intelligence reporting, such as that available in set $IR$, which closely follows the IPP narrative structure.

During fine-tuning, we divide the training set in a 35\% train and test split. We use batch size 16 and learning rate 0.0001, trained for 12 hours. The output of the model is a conditional probability of each word in the target text given the input and the previously generated words. We report a perplexity value (the exponential of the cross-entropy loss) of 11.14.

\subsection{Prefix-tuning with Narrative Prompt Sets}

Traditionally, an autoregressive decoder like GPT-Neo requires a sentence based prompt to initiate generation. To modify this requirement and to enable prompting using a narrative prompt set for intelligence report $Y$ generation, we require data-to-text \textit{prefix-tuning} \cite{li2021prefix}.

Prefix-tuning is a lightweight supplement to the fine-tuned GPT-Neo. This method keeps the GPT-Neo model parameters frozen and optimizes a sequence of continuous task-specific \textit{virtual vectors} to the key and value matrices. When the tuning process is complete, the virtual tokens are stored in a lookup table, used during inference. We use the article to plot point mappings generated by the NPCE, described in Section \ref{section:plotextraction} and the HuggingFace Parameter-Efficient Fine-Tuning  (PEFT) module for prefix-tuning. We use a beam decoding scheme and observed that adding more keywords provides increased supervision to the model and narrows the distribution of keyword context in the entire training dataset, leading to more accurate generation. More information on our evaluation can be found in Section \ref{sec:eval}.

\subsection{Large Language Model Prompting using \AlgName's EPG}
\label{section:rag}

A \AlgName generated intelligence report $Y$ on event $e$ should contain {reliable and consistent} information. $Y$ should only contain plot points that are relevant to the input query event $e$, and should exclude non-event related details. Achieving this criteria is not plausible by solely utilizing non-deterministic generative LLMs such as GPT-Neo, which are prone to \textit{output hallucinations} \cite{shuster2021retrieval}. We combat hallucinations and fulfill the above criteria by using a RAG-based approach by retrieving event narrative plot points stored in the EPG (Section \ref{section:plotextraction}) to control the prefix-tuned LLM generated output report. Our RAG approach is displayed in Figure \ref{fig:rag}.

\input{evaltable3}
%The process flow for our approach is displayed in Figure \ref{fig:rag}. \hl{instead of words, use math} 
To create a narrative prompt set for the intelligence analyst's event query $e$, \AlgName utilizes SPARQL Protocol and RDF Query Language (SPARQL) templates. These queries can be executed on the EPG to retrieve a set of plot points related to event query $e$. For example, the SPARQL query to output the \textit{Lead} narrative plot points for the event query $e = $ "\textit{Oceangate}" has been shown in Listing \ref{lst:sparql}. 
%\begin{lstlisting}[captionpos=b, caption={SPARQL generated narrative prompt set for Listing \ref{lst:sparql}.}, label=lst:sparql2,
%   basicstyle=\ttfamily,frame=single]
%\end{lstlisting}
%An intelligence analyst will utilize the IPP narrative class structure to create a series of SPARQL queries that extract information corresponding to every level of the IPP. 
%This set of queries will form a \textit{template} that will be utilized to gather and organize plot points required to construct an end to end narrative in the final generated $Y$. The template therefore contains sets of queries that are ordered by their IPP class. The head queries are executed first, followed by the body queries, and completing with the tail queries.  Each query set will extract a corresponding plot point set. 
%Each plot point in the set is forms a linearized prompt that is used as input to GPT-Neo, which generates a final report $Y$ from the set of controlled plot points. The prompts will be able to retrieve relevant plot information from the EPG to guide generation. This ensures that we utilize the \textit{entire} graph structure, including its encoded narrative features, during generation. 
%\textbf{need to figure out exactly where to put this small part.}
%Similarly, 
\AlgName includes a set of SPARQL query templates that can be leveraged to build a \textit{narrative prompt set} for the intelligence analyst query $e$. This prompt set serves as the input to the prefix-tuned GPT-Neo that outputs the intelligence report $Y$. In the next section, we describe evaluation of the generated intelligence reports. Excerpts from an intelligence report generated by \AlgName are available in Table \ref{tab:generatedtextsamples}.

\begin{lstlisting}[captionpos=b, caption={SPARQL query to retrieve Lead plot points for $e = $ ``\textit{Oceangate}'' and corresponding output}, label=lst:sparql,
   basicstyle=\ttfamily,frame=single]
SELECT Distinct ?x
WHERE {
   ?x rdf:type narr:Who. ?x rdf:type narr:What.
   ?x rdf:type narr:When. ?x rdf:type narr:Where.
   ?x rdf:type narr:Why. ?y narr:hasPlotPoint ?x.
   ?y rdf:type narr:NewsArticle.
   ?z narr:ArticleHeadline ?y.
   FILTER regex(str(?z), "Oceangate").}
Output: 
<OceanGate Expeditions, Stockton Rush, Paul-Henri 
Nargeloe, Hamish Harding, Shahzada Dawood, Suleman 
Dawood, Titanic, wreck, submersible, 18 June, 
370 miles, Newfoundland, Canada, Atlantic Ocean>
\end{lstlisting}

\section{Experimentation \& Evaluation}
\label{sec:eval}

Our evaluation study is composed of two experimental approaches. First, we automatically evaluate the semantic and syntactic quality of the generated reports using Recall-Oriented Understudy for Gisting Evaluation (ROUGE) \cite{lin2004rouge}. %, a set of metrics used for evaluating automatic summarization \cite{lin2004rouge}. 
Second, we qualitatively evaluate the reports through a human evaluation done by a group of 3 analysts. 

Table \ref{tab:generatedtextsamples}, provides samples of the generated intelligence reports for the \textit{2023 Oceangate Submersible Disaster} and the \textit{2023 Ohio Train Derailment Disaster} events. Column one displays the narrative prompt templates that were used for EPG plot point instance extraction (See Section \ref{section:rag}). Each IPP class has an associated derived narrative prompt set (Column 2), which is converted into a linearized prompt of keywords, used as input to GPT-Neo for guided text generation (See Section \ref{section:rag}). The bolded text in the generated samples (Table \ref{tab:generatedtextsamples}, Column 3) represents occurrences of the prompt keywords present in the final generated intelligence report. We limit the generation to 500 words to avoid model hallucination and inclusion of non-event related information. We found prompting with longer sequences of keywords (such as those extracted from the Lead and Body templates) resulted in more tightly coherent and semantically relevant generations, versus the Tail template, which mostly only included the extracted persuasion tactic as a prompt. We found that for instances such as these, the model would deflect from the event and sometimes include non-relevant information. Therefore, for tail generation, we limit the output to 100 words.

%Before we discuss each evaluation study set up, results, and analysis, the next section describes a qualitative analysis on the ability of \AlgName to generate intelligence reports that contain primary IPP narrative components surrounding an event. 

\subsection{Quantitative Evaluation}
%\todo{flow}
%\subsection{Automatic Evaluation}

We utilize \textit{Rouge scores} to quantitatively evaluate the efficacy of our system in generating syntactically accurate intelligence reports \cite{lin2004rouge}. %Intelligence reports can be defined as condensed accounts of the primary plot points in a complete text resource.
Rouge-$n$ in particular, allows us to compute the ratios of overlapping $n$-grams between generated reports and reference text. % Wikipedia event descriptions.
%In our evaluation approach, we model the generated intelligence reports as summaries that contain constructed narratives with a set of plot points extracted from disparately sourced articles in our corpus, $D + IR$. 
%Our goal is to quantify the ability of \AlgName to leverage the IPP features to generate intelligence reports that contain the most relevant topics (plot points) which are delineated in a logical order, similar to the orderings present in \textit{abstractive event summaries}. 
 %This requires the fine-tuned GPT-Neo to generate \textit{unique sentences}, which are not present in the original articles. 
 %\AlgName uses comma delimited extracted EPG instances as narrative prompt sets for report generation.
In particular, we use event descriptions extracted from Wikipedia as a reference set to calculate the syntactic overlap between these event descriptions and \AlgName's generated intelligence reports. The Wikipedia event descriptions are derived based on content from a variety of online sources and reporters, written by human volunteers and overseen by Wikipedia moderators. These Wikipedia reference descriptions help provide us with a lateral publicly available comparison for \AlgName's fragmented narrative construction. We use the Wikipedia Python library summaries endpoint \cite{wiki} to extract event descriptors for 50 different public events we randomly selected from $D+IR$ and calculate Rouge-$1$ and Rouge-$2$. This provides term-based measures to quantify topic-level semantic relevance and syntactic quality \cite{lin2004rouge}. Our results are displayed in Table \ref{tab:overalevall}. Rouge-$1$ refers to overlap of unigrams between \AlgName's reports and Wikipedia's event descriptors while Rouge-$2$ refers to the overlap of bigrams. 

%According to the literature, Rouge-$1$ metrics above \textit{.50} and Rouge - $2$ metric above \textit{.20} are considered plausible results \cite{schluter2017limits}.

%Wikipedia is a Python library that makes it easy to access and parse data from Wikipedia.

%Automated Abstractive summarization techniques generate new sentences while preserving themes of the original reference text, whereas extractive techniques select and combine existing sentences from a reference text to generate a summary \cite{el2021automatic}. \AlgName uses comma delimited extracted EPG instances as prompts for report generation. This reqires the fine-tuned GPT-Neo to form \textit{unique sentences}, which are not present in the original articles. \textbf{wikipedia summaries} to generate reference abstractive summaries of a test-set of news articles and use them as a reference set to compare \AlgName generated reports against. We use Rouge scores for performance and evaluation metrics. Rouge-$n$ in particular, allows us to compute the ratios of overlapping $n$-grams between generated and real text, providing term-based measures to quantify topic-level semantic relevance and syntactic quality \cite{lin2004rouge}.

\subsection{Human Evaluation (Qualitative) Study}

After evaluating the general efficacy of our model using quantitative metrics, we also conduct a human evaluation study to validate \AlgName's capability required specifically for the \textit{intelligence report generation} task. Given the high cost of this evaluation, we task a group of 3 analysts to score reports across 5 randomly selected events with two aspects: \textit{factual correctness} and \textit{language fluency}. The first criterion evaluates how well the generated report conveyed the overall narrative of the event. The second criterion evaluates grammatically correctness and fluency of the generated intelligence report. 

The analysts were given a set of 5 articles per event (total 5 events), and were tasked to manually create a \textit{single} report to convey the critical aspects across the set of 5 articles, for each separate event. This helped the analysts understand the narrative details for each of the 5 events. 
We then tasked the analysts to recommend IPP plot points from each of the 5 events. We compute Cohen's kappa \cite{mchugh2012interrater} to measure inter-annotator agreement for each of the recommended IPP plot points, keeping only the plot points that scored higher than 0.6. This helped us derive a gold standard set of 78 plot points that analysts want in the generated report, referred to as set $Gold$. We then identify the number of IPP plot points in the \AlgName's generated reports for the 5 events. The number of IPP plot points in the generated reports overlapping with $Gold$ is called support (denoted as \#Supp). The number of missing plot points that were in $Gold$ and not present in \AlgName's generated reports are called contradicting plot points (denoted as \#Cont). The average scores are displayed in Table \ref{tab:overalevall}. These are computed against the $Gold$ average of 15.6.

\begin{table}[ht]
\centering
\begin{tabular}{|clc|}
\hline
\multicolumn{3}{|c|}{\textbf{Quantitative Results}} \\ \hline
\multicolumn{1}{|c|}{\textbf{Rouge-$1$}} & \multicolumn{2}{c|}{\textbf{Rouge-$2$}} \\ \hline
\multicolumn{1}{|c|}{61.27} & \multicolumn{2}{c|}{24.51} \\ \hline
\multicolumn{3}{|c|}{\textbf{Qualitative Results}} \\ \hline
\multicolumn{1}{|c|}{\textbf{\#Supp}} & \multicolumn{1}{l|}{\textbf{\#Cont}} & \textbf{Fluency} \\ \hline
\multicolumn{1}{|c|}{13.2} & \multicolumn{1}{l|}{2.4} & 4.2 \\ \hline
\end{tabular}
\vspace{2mm}
\label{tab:overalevall}
\caption{Quantitative Qualitative Results for Generated Intelligence Reports.}
\vspace{-5mm}
\end{table}
%\todo{add gold for that in text}
To score grammatical correctness and linguistic \textit{fluency}, we adopt a 5-point Likert scale \cite{allen2007likert} ranging from 1-point (``Unacceptable'') to 5-point (``Very Acceptable'') tasking analysts to score the 5 \AlgName generated reports. The \textit{Fluency} column in Table \ref{tab:overalevall} reports averaged score (4.2/5) from 5 human analysts over the 5 generated reports.

%\todo{need to add a sentence describing table.}

\section{Conclusion \& Future Work}

Deriving narratives about events using disparately sourced information is a challenging task for an intelligence analyst. Analysts heavily rely on traditional, \textit{manual} techniques to parse large amounts of noisy OSINT data to create cohesive intelligence reports. These manual methods do not provide complete situational awareness and are prone to information gaps and inaccurate representations of dynamic events.%, and inability to separate facts from potential misinformation.
In this paper, we have described our framework \AlgName(Figure \ref{fig:sysarch}), that integrates real-time news event retrieval, narrative schema-based information extraction and representation of event concept information, and retrieval-augmented generation (RAG) of intelligence reports.

We evaluate the generated reports using quantitative Rouge evaluation metrics and through a qualitative human evaluation study. Our results show that the plot points constructed within the generated intelligence report have high semantic relevance, high coherency, and low data redundancy. In planned future work, we are exploring methods to train transformer based language models to automatically learn the structure of a variety of plot models. It is a non-uniform process to identify narrative features in natural language. We are pursuing strategies to transfer the classified plot relationships to broader events and domains.

%An analyst can use \AlgName to automatically construct intelligence reports utilizing an Event Plot Graph (EPG) to prompt a Large Language Model (LLM). Our EPG stores event narrative plot points according to the Inverted Plot Pyramid (IPP) narrative structure, specifically developed for communicating news stories. 

%We utilize narrative prompt engineering and a Retrieval Augmented Generation (RAG) approach to fine-tune the generative LLM GPT-Neo, to output an intelligence report. % with respect to event time, setting, and theme. 

\bibliographystyle{unsrt}
\bibliography{references}

\end{document}

%% file: plottable.tex
% \usepackage{multirow}
% \usepackage{multirow}
\begin{table*}[]
\footnotesize
\centering
\begin{tabular}{|c|l|l|}
\hline
\textbf{Inverted Pyramid Class} & \multicolumn{1}{c|}{\textbf{Plot Element}} & \multicolumn{1}{c|}{\textbf{Notation}} \\ \hline
\multirow{5}{*}{\begin{tabular}[c]{@{}c@{}}Lead: Summary of the most noteworthy \\ details and main story objective/event.\end{tabular}} & Who: Identification of the subject or persons involved. & $p_{who}$ \\ \cline{2-3} 
 & What: Occurrences of scenes, incidents, artifacts, or actions. & $p_{what}$ \\ \cline{2-3} 
 & When: Recorded timestamps and dates. & $p_{when}$ \\ \cline{2-3} 
 & Where: Geographic regions and locations mentioned. & $p_{where}$ \\ \cline{2-3} 
 & Why: The cause and reason to describe event occurence. & $p_{why}$ \\ \hline
\multirow{5}{*}{\begin{tabular}[c]{@{}c@{}}Body: Series of crucial delineated incidents \\ that build main story objective/event.\end{tabular}} & Evidence: Supporting details surrounding an event. & $p_{evidence}$ \\ \cline{2-3} 
 & Quotes: Phrases noted by involved persons. & $p_{quote}$ \\ \cline{2-3} 
 & Media (Photos): Digital Image Object (HTML DOM image element) & $p_{photo}$ \\ \cline{2-3} 
 & Media (Video): Recorded Video Object (HTML DOM video element) & $p_{video}$ \\ \cline{2-3} 
 & Media (Audio): Recorded Audio Object (HTML Dom audio Element.) & $p_{audio}$ \\ \hline
\multicolumn{1}{|l|}{\multirow{3}{*}{\begin{tabular}[c]{@{}l@{}}Tail: Conclusory remarks \\ and journalistic opinions.\end{tabular}}} & Journalistic Opinion: Non-fact based judgements about event. & $p_{opinion}$ \\ \cline{2-3} 
\multicolumn{1}{|l|}{} & Persuasion Tactic: Instances of rhetorical dimensions (ethos, pathos, logos) & $p_{tactic}$ \\ \cline{2-3} 
\multicolumn{1}{|l|}{} & Sentiment: Emotional tone and affective state information. & $p_{sentiment}$ \\ \hline
\end{tabular}
\vspace{2mm}
\caption{Textual features representing plot elements for each Inverted Pyramid Plot Model class.}
\label{tab:ipp}
\vspace{-5mm}
\end{table*}

%\todo{table competion alignment caption}

%% file: evaltable3.tex
\begin{table*}[ht]
\footnotesize
\centering
\begin{tabular}{|lll|}
\hline
\multicolumn{3}{|c|}{\textbf{2023 Oceangate Submersible Disaster}} \\ \hline
\multicolumn{1}{|c|}{\textbf{Prompt Template}} & \multicolumn{1}{c|}{\textbf{Prompt}} & \multicolumn{1}{c|}{\textbf{Generated Text}} \\ \hline
\multicolumn{1}{|l|}{\begin{tabular}[c]{@{}l@{}}Lead Template\\ \textless{}$p_{who}$, $p_{what}$, \\ $p_{when}$, $p_{where}$,\textgreater{}\end{tabular}} & \multicolumn{1}{l|}{\begin{tabular}[c]{@{}l@{}}\textless{}OceanGate Expeditions, Stockton Rush, \\ Paul-Henri Nargeloe, Hamish Harding, \\ Shahzada Dawood, Suleman Dawood, \\ Titanic, wreck, site, submersible, \\ 18 June, 370 miles, Newfoundland, \\ Canada, 1 hour and 45 minutes, \\ Atlantic Ocean\textgreater{}\end{tabular}} & \begin{tabular}[c]{@{}l@{}}The \textbf{OceanGate Expeditions} submarine is the subject of the \\ rescue operation for \textbf{Hamish Harding},  \textbf{Stockton Rush}, \\ \textbf{Paul-Henri Nargeloe}, \textbf{Shahzada Dawood}, \textbf{Suleman} \\ \textbf{Dawood} were on board for an expedition into the \textbf{Titanic}\\ \textbf{wreck site}. The Titan \textbf{submersible} went missing \textbf{18 June} about \textbf{370} \\ \textbf{miles} off the coast of \textbf{Newfoundland} in \textbf{Canada}. The \\ submersible lost contact just \textbf{1 hour and 45 minutes} after diving \\ into the \textbf{Atlantic Ocean}.\end{tabular} \\ \hline
\multicolumn{1}{|l|}{\begin{tabular}[c]{@{}l@{}}Body Template\\ \textless{}$p_{evidence}$, \\ $p_{quote}$,\textgreater{}\end{tabular}} & \multicolumn{1}{l|}{\begin{tabular}[c]{@{}l@{}}\textless{}debris, ROV, seafloor, titanic, sonar, \\ banging, noises, implosion, injury, dead, \\ safety, regulations, failure, United States \\ Coast Guard said “likely died from the \\ failure of the craft”\textgreater{}\end{tabular}} & \begin{tabular}[c]{@{}l@{}}\textbf{Debris} was found by a \textbf{ROV} near the \textbf{Titanic} wreck. \textbf{Sonar} detects \\ \textbf{banging} sounds from underneath the water. The banging sounds \\ indicate an unfortunate \textbf{implosion}.  \textbf{The United States Coast} \\ \textbf{Guard} claims the crew and passengers died from the implosion.\end{tabular} \\ \hline
\multicolumn{1}{|l|}{\begin{tabular}[c]{@{}l@{}}Tail Template\\ \textless{}$p_{tactic}$,\textgreater{}\end{tabular}} & \multicolumn{1}{l|}{\begin{tabular}[c]{@{}l@{}}\textless{}overshadowed the bigger Greece \\ migrant vessel disaster\textgreater{}\end{tabular}} & Coverage \textbf{overshadowed} the greater Mediterranean disaster. \\ \hline
\multicolumn{3}{|c|}{\textbf{2023 Ohio Train Derailment Disaster}} \\ \hline
\multicolumn{1}{|c|}{\textbf{Prompt Template}} & \multicolumn{1}{c|}{\textbf{Prompt}} & \multicolumn{1}{c|}{\textbf{Generated Text}} \\ \hline
\multicolumn{1}{|l|}{\begin{tabular}[c]{@{}l@{}}Lead Template\\ \textless{}$p_{who}$, $p_{what}$, \\ $p_{when}$, $p_{where}$,\textgreater{}\end{tabular}} & \multicolumn{1}{l|}{\begin{tabular}[c]{@{}l@{}}\textless{}Norfolk Southern Train, chemicals, \\ East Palestine, Ohio, 9pm, 3 February\textgreater{}\end{tabular}} & \begin{tabular}[c]{@{}l@{}}\textbf{Norfolk Southern Train} carrying hazardous \textbf{chemicals} suffered \\ a mechanical failure. The derailment took place in the village \\ of \textbf{East Palestine, Ohio} at \textbf{9pm }on \textbf{3 February}.\end{tabular} \\ \hline
\multicolumn{1}{|l|}{\begin{tabular}[c]{@{}l@{}}Body Template\\ \textless{}$p_{evidence}$, \\ $p_{quote}$,\textgreater{}\end{tabular}} & \multicolumn{1}{l|}{\begin{tabular}[c]{@{}l@{}}\textless{}derailment, cars, EPA, hazardous, \\ chemicals, contaminants, cancer, \\ gas, safety, The EPA said it “did not \\ detect chemical contaminants at concerning\\  levels in the hours after venting.”\textgreater{}\end{tabular}} & \begin{tabular}[c]{@{}l@{}}The cars had \textbf{derailed}, including the cars that were carrying \\ \textbf{hazardous} materials. The \textbf{EPA} claims no concerning \\ health risks. The chemicals are linked to increased \\ risk of \textbf{cancer}.\end{tabular} \\ \hline
\multicolumn{1}{|l|}{\begin{tabular}[c]{@{}l@{}}Tail Template\\ \textless{}$p_{tactic}$,\textgreater{}\end{tabular}} & \multicolumn{1}{l|}{\begin{tabular}[c]{@{}l@{}}\textless{}overblown characterisations about \\ the derailment disaster\textgreater{}\end{tabular}} & \begin{tabular}[c]{@{}l@{}}Reporters made \textbf{overblown} accusations about the derailment \\ disaster.\end{tabular} \\ \hline
\end{tabular}
\label{tab:generatedtextsamples}
\vspace{2mm}
\caption{Guided Intelligence Report generation using IPP prompt templates.}
\vspace{-7mm}
\end{table*}